# Assessing the Role of Random Forests in Medical Image Segmentation

Dennis HARTMANN[1], Dominik MÜLLER[1], Iñaki SOTO-REY[1,2] and Frank KRAMER[1]

[1]*IT-Infrastructure for Translational Medical Research, University of Augsburg, Germany,*
[2]*Medical Data Integration Center, Institute for Digital Medicine, University Hospital Augsburg, Germany*

**Abstract.** Neural networks represent a field of research that can quickly achieve very good results in the field of medical image segmentation using a GPU. A possible way to achieve good results without GPUs are random forests. For this purpose, two random forest approaches were compared with a state-of-the-art deep convolutional neural network. To make the comparison the PhC-C2DH-U373 and the retinal imaging datasets were used. The evaluation showed that the deep convolutional neutral network achieved the best results. However, one of the random forest approaches also achieved a similar high performance. Our results indicate that random forest approaches are a good alternative to deep convolutional neural networks and, thus, allow the usage of medical image segmentation without a GPU.

**Keywords.** Medical imaging, random forest, biomedical image segmentation, supervised learning

## 1. Introduction

Automated medical image analysis is an active and highly demanded research field, especially for clinical decision support. One technique of automated medical image analysis is medical image segmentation. In this technique, image points are checked for characteristics and, thus, certain structures in an image are highlighted. However, since it is not feasible to incorporate all characteristics for all types of medical conditions into a single algorithm, various approaches have been developed. One of these approaches are deep convolutional neural networks (DCNNs). DCNNs are state-of-the-art in this research field [1]. This is because they achieved extraordinary high performance comparable to human radiologists [2, 3]. However, for processing the high volume of data, neural network models require costly high-performance GPU hardware.

Random forests (RFs) are widely popular in the scientific field of machine learning. They can be trained quickly without the need of a GPU, which makes their use in a clinical environment easier. Nevertheless, the use of RF for image segmentation is rather uncommon and, to our knowledge, RF has not been directly compared to DCNN in terms of performance and usability.

---

[1] Corresponding Author, Dominik Müller, IT Infrastructure for Translational Medical Research, Alter Postweg 101, 86159 Augsburg, Germany; E-mail: dominik.mueller@informatik.uni-augsburg.de.

In this work, we want to assess the capabilities of random forests for medical image segmentation. By implementation as well as application, we want to compare the performance and usability differences between two random forest pipelines and one deep convolutional neural network model.

**2. Methods**

*2.1. Datasets*

To enable a robust and thorough evaluation of the three algorithms, we selected the following datasets:
The first dataset was the PhC-C2DH-U373. This dataset of Dr. S. Kumar. Department of Bioengineering, University of California at Berkeley, Berkeley CA (USA) showed Glioblastoma-astrocytoma U373 cells on a polyacrylamide substrate. The aim of this dataset was to segment the cells. For the recording a Nikon microscope with a Plan Fluor DLL 20x/0.5 objective lens, a pixel size of 0.65x0.65 and a time step of 15 minutes was used. The resulting dataset consisted of 34 images. [4]
The second dataset was the retinal imaging dataset. The dataset consisted of the 19 recordings from Hoover et al. and 40 images from a Netherlands screening program. It deals with photographs of the back of the eye. The focus of this dataset was on the segmentation of the vessels. The images of the Netherlands were taken with a Canon CR5 nonmydriatic 3CCD at 45° field of view and a resolution of 768x584 pixels. The images of Hoover et al. were taken with a TopCon TRV-50 fundus camera at 35° field of view and a resolution of 700x605 pixels. The images were labeled by Michael D. Abràmoff an ophthalmologist, a computer science student and Joes Staal, which have been trained by the ophthalmologist. Of these 59 images, 20 were used. [5]

*2.2. Preprocessing and Image-Augmentation*

For reducing the complexity of the segmentation task by simplifying the pattern detecting process and to increase the imaging data, various functions have been implemented.
First the images were converted to greyscale. After that the size of the images was set to the maximum image size of the dataset for the RF approaches. For this purpose, the missing areas were filled with 0. To extract more images from the existing the python library Albumentations was used [6]. This has resulted in an enlargement of the image quantity by a factor of 10. This corresponds to an increase in the number of images in the PhC-C2DH-U373 dataset from 34 to 340 and in the Retinal Imaging Dataset from 20 to 200 images. Therefore, the images were altered in their brightness, were rotated, the gamma was changed, an elastic transform and an optical distortion were executed, the images were flipped, random rotated, compressed as well as the contrast changed. Then, the images have been resized to 512x512 pixels. Afterwards, the resized images were processed with a Sobel filter to get all data for the RF approaches.

*2.3. Random Forest Model*

The random forests were trained with 100 trees, a maximum depth of 40 and using a cross-entropy for measuring the quality of the splits. In order to consider the task of

image segmentation at different levels of complexity, the following approaches were implemented:

*2.3.1. Feature Extraction Architecture*

The feature extraction architecture (RF-FE) uses a pixel-wise training approach. Therefore, 4 features per pixel were calculated. These included the pixel-value of the original value intensity, the mean of the 13x13 pixel neighborhood of the pixel in the original image as well as of the Sobel filtered image. Additionally, a square of 13x13 pixels of the neighborhood of the pixel of the original image was added to the feature set.

*2.3.2. Whole Image Architecture*

The whole image architecture (RF-WI) utilized whole images as input for the model by simply defining all pixels as a single feature array. Therefore, the Sobel filtered image was used to train one RF model.

*2.4. Deep Convolutional Neural Network Model*

The DCNN pipeline was built using MIScnn [7]. Therefore, the U-Net model [8] with batch normalization, 100 epochs, batch size 2, whole images as feature input and the sum of the Tversky index as well as the categorical cross-entropy as loss function was used. Furthermore, we utilized an Adam optimizer with a dynamic learning rate starting from 0.001 up to 0.00001 with a 0.1 decrease every 5 epochs without loss decrease. Additionally, we applied an early stopping technique after 10 epochs without loss improvement.

**3. Results**

All algorithms were trained with 80% and tested with 20% of the data, if possible. To ensure a good comparability of the algorithms' performance, the data were split in advance. Afterwards, all algorithms and architectures were trained with the same training data and tested on the same unseen testing set. The resource consumption for the individual algorithms can be found in Table 1. In Table 2, the comparison of the algorithm's performance and on Figure 1, one prediction of each of the algorithms for one image of each dataset, are visualized. Figure 2 shows the performance measured against various evaluation metrics in comparison using boxplots and Figure 3 compares the results of the RF-FE with the DCNN via a scatter plot of the measured performance.

Table 1. Maximum hardware and time requirements for model training of the implemented algorithms.

| Algorithm | Memory | GPU | Time |
|---|---|---|---|
| **Feature Extraction** | 207 GB | X | 6.79h |
| **Whole Image** | >220 GB | X | 0.71h |
| **DCNN** | 3 GB | 7 GB | 0.43 h |

Table 2. Performance comparison through macro-averaged accuracy, Dice similarity coefficient, intersection-over-union (IoU) and sensitivity (Sens.) of the implemented algorithms.

| Algorithm | PhC-C2DH-U373 | | | | Retinal Imaging | | | |
|---|---|---|---|---|---|---|---|---|
| | Accuracy | Dice | IoU. | Sens. | Accuracy | Dice | IoU. | Sens. |
| **Feature Extraction** | 0.98 | 0.85 | 0.75 | 0.84 | 0.95 | 0.68 | 0.52 | 0.58 |
| **Whole Image** | 0.95 | 0.23 | 0.16 | 0.17 | 0.91 | 0.00 | 0.00 | 0.00 |
| **DCNN** | 0.99 | 0.90 | 0.84 | 0.89 | 0.96 | 0.77 | 0.63 | 0.75 |

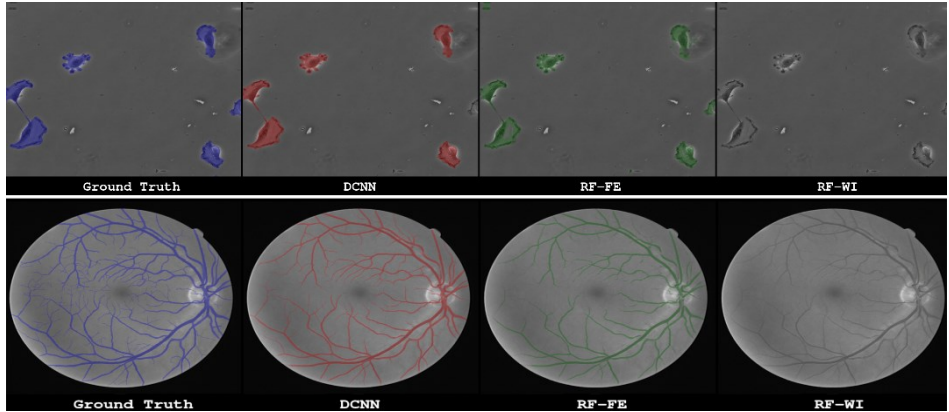

**Figure 1**. Predictions of the different algorithms compared to the ground truth for one image of each of the datasets.

### 3.1. Random Forest Results

#### 3.1.1. Feature Extraction Architecture

On the PhC-C2DH-U373 dataset an accuracy of 0.98, a Dice similarity coefficient of 0.85, an intersection-over-union of 0.75 and a sensitivity 0.84 was obtained. A macro-averaged accuracy of 0.95, a Dice similarity coefficient of 0.68, an intersection-over-union of 0.52 and a sensitivity 0.58 was achieved on the retinal imaging dataset.

#### 3.1.2. Whole Image Architecture

The Whole Image Architecture achieved a macro-averaged accuracy of 0.95, a Dice similarity coefficient of 0.23, an intersection-over-union of 0.16 and a sensitivity of 0.17 on the PhC-C2DH-U373 dataset. For the training, the dataset had to be limited to 16 images due to the RAM limitation. A macro-averaged accuracy of 0.91, a Dice similarity score of 0.00, an intersection-over-union of 0.00 and a sensitivity of 0.00 was obtained on the retinal imaging dataset.

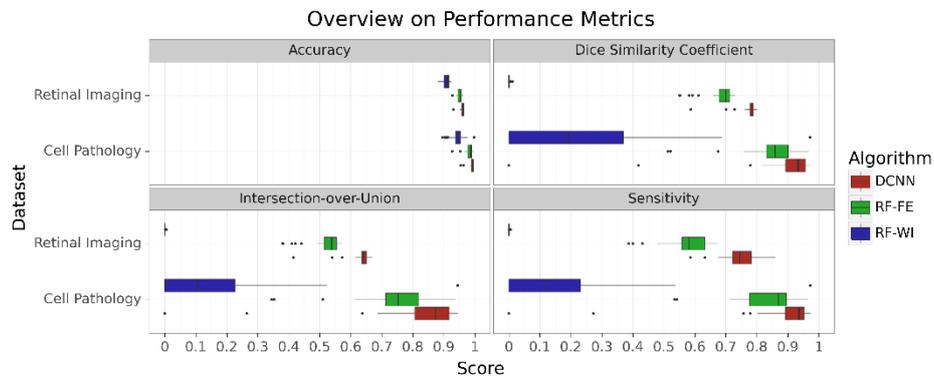

**Figure 2**. Comparison of the segmentation algorithm performances for the datasets via boxplots.

*3.2. DCNN Results*

The DCNN model obtained a macro-averaged accuracy of 0.99, a Dice similarity score of 0.90, an intersection-over-union of 0.84 and a sensitivity of 0.89 on the PhC-C2DH-U373 dataset. On retinal imaging dataset a macro-averaged accuracy of 0.96, a Dice similarity score of 0.77, an intersection-over-union of 0.63 and a sensitivity of 0.75 was achieved.

**4. Discussion**

In terms of general accuracy and sensitivity, our evaluation showed that the DCNN model achieved superior results than the RF approaches. It also required significantly less memory. Nevertheless, the feature extraction architecture achieved accurate and precise results, which were comparable with the DCNN performance. The whole image architecture, on the other hand, achieved only poor results. The better results of the DCNN could be due to the fact that neural network architectures allow more efficient recognition of areas of interest and have an improved ability to recognize general patterns in the images. However, since GPUs are not always available, a RF with a feature extraction architecture could be used as an alternative. By requiring only sufficient CPU and RAM hardware, such as commonly found in hospitals, RFs allow more flexible access on precise medical image segmentation for clinical decision support. However, a large amount of hardware memory is necessary for this in order to achieve optimal results.

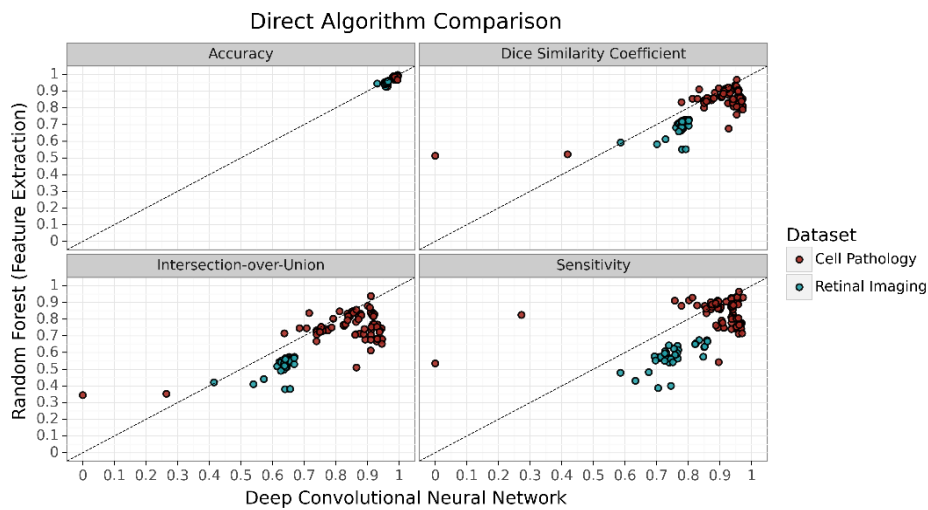

**Figure 3**. Direct performance comparison of the segmentation algorithms through multiple evaluation metrics via a scatter plot. The gray dashed line indicates the area for equal performance.

Further improvements to the feature extraction architecture can be made by adding more features like the pixel values from Sobel filtering, additional Z-Score normalization on image intensity values as well as their neighborhood mean. Various recent studies achieved excellent results on their own for RF medical image segmentation and introduced new features as well as methods that could further enhance the results of our implementation for universal medical conditions. As applications, Baka et al. demonstrated accurate bone segmentation in ultrasound imaging with a recall of 0.86 at a precision of 0.82 [9], Chen et al. achieved state-of-the-art results for brain lesion segmentation on the popular BRATS dataset (2015 & 2018) comparable to DCNN approaches [10] and Cao et al. proved the usability and efficiency of RFs for automated glomerular basement membrane segmentation in transmission electron microscopy imaging [11].

## 5. Conclusions

In this paper, two random forest architectures for medical image segmentation were presented. As a first approach, a whole image random forest architecture was implemented utilizing complete images as feature array. Additionally, a feature extraction architecture for random forests were added which analyzes each pixel individually via multiple features. The two approaches were compared with a state-of-the-art deep convolutional neural network model based on a U-net architecture.

In our analysis, we demonstrated that deep convolutional neural networks as well as random forests based on a feature extraction architecture were able to achieve excellent results for medial image segmentation. Still, the whole image architecture achieved no reliable results at all for more complex segmentation tasks like vessel extraction in retinal imaging.

However, the random forest approaches have the advantage that they can also be used in smaller IT infrastructures without the need of a GPU cluster. On the other hand, a large amount of hardware memory is required for optimal segmentation precision.

## 6. Contributions of the Authors

Dr. Frank Kramer and Dr. Iñaki Soto-Rey contributed to the coordination, review, and correction of the manuscript. Dominik Müller was in charge of the coordination, study design conception, evaluation and manuscript revision. Dennis Hartmann was in charge for the implementation, data analysis and interpretation as well as for manuscript drafting and revision.

## 7. Conflict of Interest

None declared.

## 8. Funding

This work is a part of the DIFUTURE project funded by the German Ministry of Education and Research (Bundesministerium für Bildung und Forschung, BMBF) grant FKZ01ZZ1804E.